\documentclass[twocolumn,prl,ams,superscriptaddress,floatfix]{revtex4}
\usepackage{graphicx}
\usepackage{epstopdf}
\usepackage{dcolumn}
\usepackage{amssymb,amsmath}
\usepackage{bm}

\begin{document}
\title{Developed Smectics: When Exact Solutions Agree}
\author{Gareth P. Alexander}
\affiliation{Centre for Complexity Science, Zeeman Building, University of Warwick, Coventry, CV4 7AL, UK}
\author{Randall D. Kamien}  
\affiliation{Department of Physics and Astronomy, University of Pennsylvania, Philadelphia, PA 19104-6396, USA}
\author{Christian D. Santangelo}
\affiliation{Department of Physics, University of Massachusetts, Amherst MA, 01003, USA}
\date{\today}

\begin{abstract}
In the limit where the bending modulus vanishes, we construct layer configurations with arbitrary dislocation textures by exploiting a connection between uniformly-spaced layers in two dimensions and developable surfaces in three dimensions. We then show how these focal textures can be used to construct layer configurations with finite bending modulus.
\end{abstract}
\maketitle

When subject to frustrating boundary conditions or extreme strains, liquid crystals, superfluids, and magnets will locally rise into their higher-symmetry phases resulting in point, line, and planar defects \cite{Klemanbook}.  Energetic considerations determine the dimensionality of these defects;  in some systems, rigorous results demonstrate that the energy minimizers will have point or line defects \cite{Uhlenbeck,Kinderlehrer}.  Smectic liquid crystals represent a special challenge as they are described by an essentially nonlinear elasticity theory \cite{Klemanbook,pleiner,bk,marchenko} that gives rise to anomalous elasticity \cite{GP}, dynamics \cite{TMR,Miln}, and qualitatively modified ground states \cite{brener,ishikawa}. 
These nonlinearities are generic features of elastic systems with free surfaces \cite{wittenRMP} and, thus, smectics are ideal systems for understanding elastic geometric nonlinearities in general.
Previously, we have studied smectic liquid crystals in the limit where the bending energy is neglected so that the layer spacing is strictly constant \cite{columns,prl}.  Here we extend some of these techniques by employing a connection between developable surfaces in three dimensions and uniformly-spaced layers in two dimensions, allowing us to find layer configurations for any specified dislocation texture.  We compare these solutions with exact solutions to the nonlinear elasticity \cite{brener} equations which only allow superposition of dislocations along a single line \cite{santangelo03,newell}.  
Not only do the two solution methods agree, but the geometric construction explains the fundamental asymmetry of the smectic strain field around a dislocation, first predicted by Brener and Marchenko \cite{brener}, and sheds light on the simple, topologically based, Bogomol'nyi-Prasad-Sommerfield (BPS) bound \cite{santangelo03,prsa}. We exploit this understanding to construct textures for dislocations separated by a finite number of layers with finite bending rigidity.

The order in a smectic is characterized by the phase field $\phi({\bf x})$ appearing in the density modulation $\delta\rho \propto \cos[2\pi\phi({\bf x})/a]$, where $a$ is the natural layer spacing. 
In terms of $\phi$ the free energy is the sum of compression and bending contributions
\begin{equation}\label{eq:f}
F = \frac{B}{2} \int d^2x\, \left\{\left(\vert{\bm\nabla}\phi\vert -1\right)^2 + \lambda^2 \left({\bm\nabla}\cdot\frac{{\bm\nabla} \phi}{|{\bm\nabla} \phi|}\right)^2\right\} ,
\end{equation}
where $B$ is the compression modulus,  $\lambda=\sqrt{K_1/B}$ is the penetration length and $K_1$ is the bending modulus.  In smectics A, the normal to the smectic layers is the nematic director ${\bf n}=\nabla\phi/\vert\nabla\phi\vert$. Geometrical and topological insight is gained by considering the surface $\left[x,y,\phi(x,y)\right]\in\mathbb{R}^3$ with surface normal ${\bf N}=\left[-\partial_x\phi,-\partial_y\phi,1\right]/\sqrt{1+\vert{\bm\nabla}\phi\vert^2}$ \cite{pnas}. Here, we shall focus our attention on the limit $\lambda\ll a$, or $K_1\rightarrow 0$, where bending becomes unimportant compared to compression. Indeed, when $\lambda=0$ the free-energy is strictly minimized when $\vert{\bm\nabla}\phi\vert=1$; 
differentiating $\left({\bm\nabla}\phi\right)^2=1$, we have:
\begin{equation}
\left(
\begin{array}{ll}
\partial_x^2\phi & \partial_x\partial_y\phi\\
\partial_y\partial_x\phi & \partial_y^2\phi
\end{array}
\right)\left(
\begin{matrix}
\partial_x\phi \\ \partial_y\phi
\end{matrix}\right) = 0 ,
\end{equation}
which requires the Gaussian curvature, $K\propto \partial_x^2\phi\partial_y^2\phi - \left(\partial_x\partial_y\phi\right)^2=0$.  
It follows from Gau\ss's {\sl Theorem Egregium} that our surface must be isometric to the plane, so it can be built out of sections of planes, cones, cylinders, and tangent-developable surfaces.  The constant angle condition further restricts to planes, cones, and the development of cylindrical helices \cite{nistor}.   

It is amusing that the latter can be used to generate uniformly-spaced involutes of curves \cite{bouligand,columns2}; though known to the ancients \cite{Eisenhart}, we will briefly review the connection between level sets of constant-angle, developable surfaces and involutes. Consider a curve ${\bf R}(\sigma)=[x(\sigma),y(\sigma),z(\sigma)]$ in $\mathbb{R}^3$, parameterized by its arclength $\sigma$, with Frenet-Serret frame $[{\bm t},{\bm \nu},{\bm \beta}]=[\dot{\bf R},\dot{\bm t}/\kappa,{\bm t}\times{\bm \nu}]$, curvature $\kappa(\sigma)> 0$ and torsion $\tau$.  The tangent developable surface is defined in terms of the curve and its family of tangents: ${\bf X}(\sigma_1,\sigma_2) = {\bf R}(\sigma_1) - \sigma_2{\bm t}(\sigma_1)$ for $\sigma_2\ge 0$. Note that the unit normal to the surface ${\bf N}(\sigma_1,\sigma_2) = \partial_1{\bf X}\times\partial_2{\bf X}/\vert \partial_1{\bf X}\times\partial_2{\bf X}\vert = {\bm\beta}(\sigma_1)$, the curve's binormal at $\sigma_1$. It follows that ${\bf N}$ only depends on $\sigma_1$ and so the Gau\ss\ curvature vanishes.  If the angle between $\bf N$ and $\hat{\bf z}$ is constant then so is the angle between $\bm\beta$ and $\hat{\bf z}$.  Differentiating with repect to $\sigma$, we have 
$0={\hat{\bf z}}\cdot\dot{\bm\beta}= -\tau{\hat{\bf z}}\cdot{\bm\nu}$ so ${\bm\nu}$ lies in the $xy$-plane. Define the surface curve ${\bm\gamma}(s) = {\bf X}(\sigma_1+s,\sigma_2 +s)$ with tangent $\dot{\bm\gamma}(s)=  (\sigma_2+s)\kappa(\sigma_1+s){\bm\nu}(\sigma_1+s)$. $\bm\gamma$ 
lies in a plane of constant $z=c$ and ${\bm\gamma}(s)$ sweeps out an involute starting at $s=0$ on the planar curve ${\bf R}_\perp(s) \equiv [x(\sigma_1+s),y(\sigma_1+s),c]$. 
Apart from concentric circles, any set of uniformly-spaced involutes will generate an evolute curve which constitutes a singularity or edge of the surface and where the bending of the involutes diverges. Since this will generate a two-dimensional region without smectic order, we will not consider such cases, although surfaces like this are liable to play a role in sample cells with large inclusions.   Here we are interested in defects that can be reduced to points and lines and so we only consider constant angle cones and planes.  For convenience we set the constant angle to be $\pi/4$.

\begin{figure}[t]
\centering
\includegraphics[width=6cm]{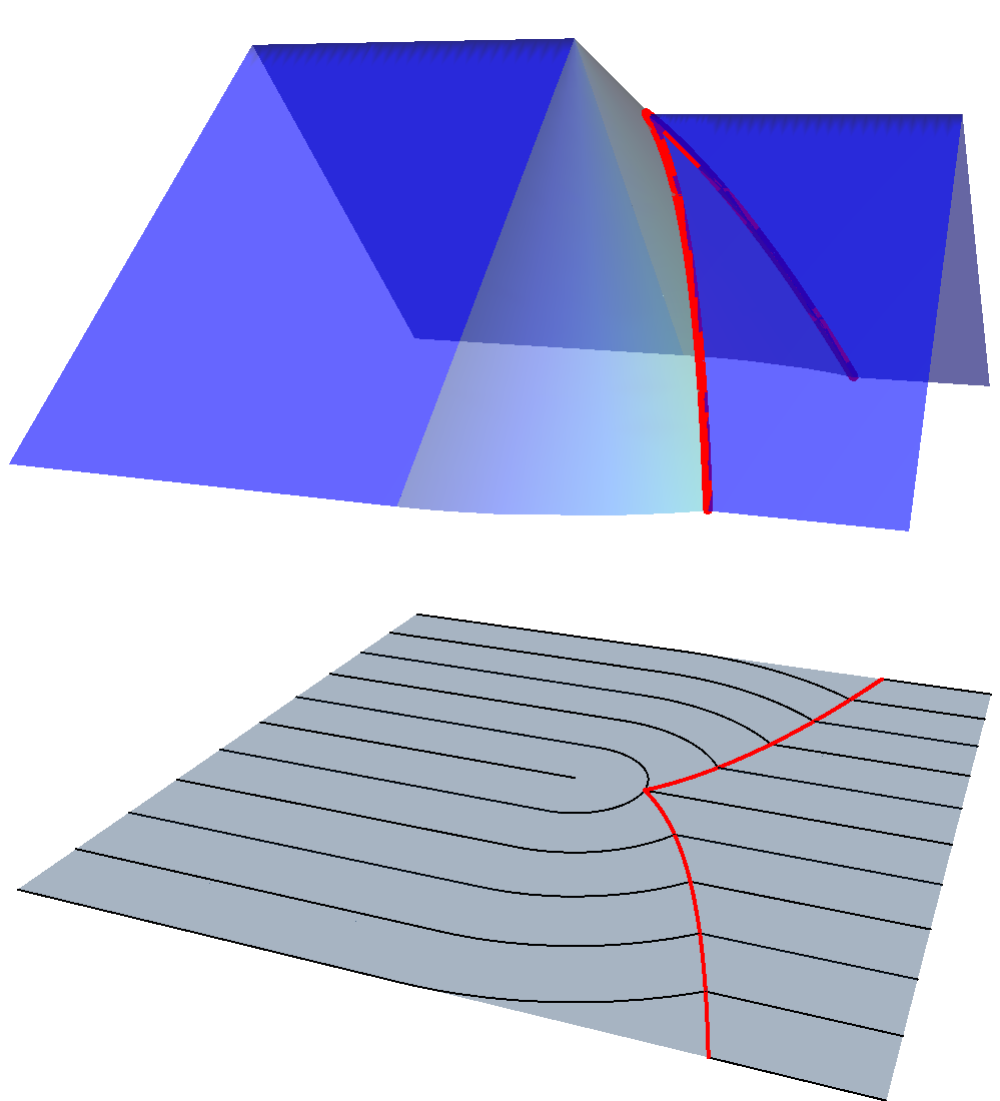}
\caption{(Color online) We construct the two-dimensional layers by taking level sets of a piecewise developable (Gaussian curvature $K=0$) surface which makes a constant angle with the $\hat{\bf z}$ direction. From back left to close right, the surface is made of two intersecting planes which end and attach to pieces of cone which necessarily intersect another set of parallel planes on a parabola. \label{1disfocal}}
\end{figure}

Smectics enjoy two types of point defects, disclinations and dislocations.  In the language of surfaces the disclinations are critical or singular points on the graph of $\phi$. Dislocations can be constructed by choosing $\phi=x+(b/2\pi)\arg{(x+iy)}$ to be a tilted helicoid \cite{pnas}, resulting in a two-dimensional smectic with bending and compression deformations.  However, we can also build a dislocation with vanishing compression with lines across which the director jumps discontinously, thus being visible under light microscopy.  
To this end, consider the construction of an edge dislocation shown in Fig. \ref{1disfocal}. Two planes meeting along a ridge are connected to two similar planes, that meet along a ridge at a lower height ($b/2$ lower where $b \in a \mathbb{Z}$ is the Burgers scalar), by a portion of a cone. The cone's apex coincides with the endpoint of the upper ridge and the transition from plane to cone is Lipschitz $C^1$. However, the intersection with the lower pair of planes introduces a cusp, or focal curve, along which the normal changes discontinuously and the surface is only Lipschitz $C^0$, as is the director field. This focal curve consists of part of a pair of parabol\ae. Taking level sets of the surface produces an uniformly-spaced smectic texture for a dislocation.   
Aside from the point defect corresponding to the cone's vertex, there is a focal set consisting of the two parabolic segments $x^2=b|y|+b^2/4$.  Recall that in the linear theory the elastic response is concentrated in two full parabolic regions above and below the defect \cite{ntsma}.  The present construction only generates compression strain on the ``right'' side of the defect.  

\begin{figure}[t]
\centering
\includegraphics[width=.4\textwidth]{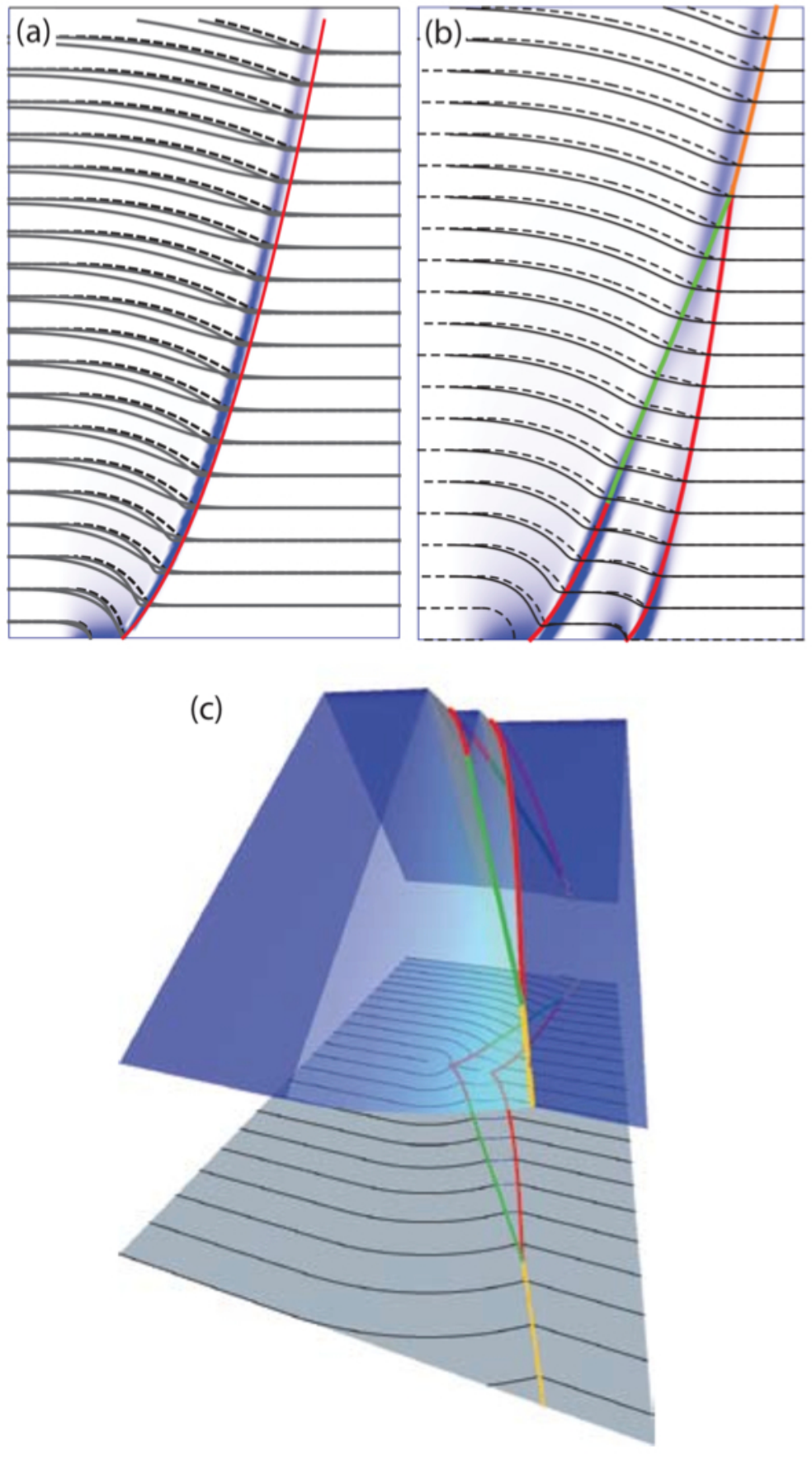}
\caption{(Color online) Comparison of focal and BPS constructions: the dashed lines show the layers from the focal construction in Fig. \ref{1disfocal} for the layers on one side of (a) one or (b) two dislocations. The solid curves are level sets of $\phi=y-u(x,y)$, where $u$ is determined by BPS evolution, starting with the first layer of the focal construction, for $\lambda=0.05$ and $\lambda=0.1$. The background is shaded according to the compression energy of the asymptotic BPS solution, Eq. (\ref{eq:1BPS}), for $\lambda=0.05$.  The parabolic focal line is $y=\pm(x^2/b - b/4)$; we use the vertical offset in the BPS solution \cite{brener,santangelo03}. (c) The focal construction for two dislocations, built by attaching two single dislocations as in Fig. \ref{1disfocal}.  Note that now some of the focal lines arise from the intersection of cones with cones and are pieces of hyperbol\ae, not parabol\ae .  
\label{fig2G}}
\end{figure}

In the presence of a defect, BPS minimizers of \eqref{eq:f} and related free energies were found \cite{brener,newell,santangelo03} and, for small $\lambda/y$ \cite{prsa}, the  displacement for a single defect at $(x,y)=(0,0)$ is $u(x,y)\equiv y - \phi(x,y)$ is
\begin{equation} \label{eq:1BPS}
u(x,y) = 2\lambda{\rm sgn}(y) \ln \left[1+\left(\text{e}^{-b/(4 \lambda)}-1\right){\rm E}\left(\frac{x}{2\sqrt{\lambda |y|}}\right) \right] ,
\end{equation}
where ${\rm E}(x) \equiv (\pi)^{-1/2}\int_{-\infty}^x dt \exp(-t^2)$ is the error function.  The associated compression strain $e$ for $y>0$ scales as
\begin{equation}
\partial_y u  = \frac{-x\sqrt{\lambda}}{2\sqrt{\pi y^3}}\frac{\left(\text{e}^{-b/(4 \lambda)}-1\right) \text{e}^{-x^2/(4\lambda y)}}{1+\left(\text{e}^{-b/(4 \lambda)}-1\right){\rm E}\left(\frac{x}{2\sqrt{\lambda y}}\right)} .
\end{equation}
For large $\lambda/b$ this reproduces the symmetric, linear strain field.  However, as $\lambda/b\rightarrow 0$, we have $\partial_y u \sim \theta(x) \delta(y - x^2/b)$, half of a parabola on the side with {\sl fewer} layers, and, as shown in Fig. \ref{fig2G}, in 
agreement with the focal construction.    Though the shape of the parabola is identical in the focal and BPS solutions, we note that there is a vertical offset of $b/4$ between them.  Because the BPS solution is based only on a step-function boundary condition at $y=0$ used to satisfy the topology at infinity, we do not expect the near-defect details to be reproduced, but for large $x$ and $y$, the solutions agree as shown in \cite{prsa}.  

Why should the strain be  asymmetric \cite{brener}? Recall that the nonlinear compression strain $e$ measures the deviation of the wavenumber  $q=2\pi/d$ from $q_0=2\pi/a$, $e\propto (q-q_0)^2$ and so, away from the linear regime, compression $d<a$ is more energetic than dilation $d>a$.  It follows that in the equal-spacing limit, the texture will preferentially distort on the dilated side.
The presence of a focal line in the BPS solution also is not a mystery.  Differentiating the BPS equation 
\begin{equation}\label{eq:BPSevolution}
\partial_y u - \frac{1}{2} \bigl( \partial_x u \bigr)^2 = \lambda \partial_x^2 u ,
\end{equation}
with respect to $x$ yields the Burgers equation $\partial_yv-v\partial_xv=\lambda\partial_x^2v$ for $v=\partial_xu$. As is well known, the inviscid Burgers equation has straight characteristics and produces asymptotically parabolic shock curves as we have here \cite{Whitham}.  In comparison, the focal construction arises from constructing characteristics of the geodesic condition $\left({\bf n}\cdot{\bm\nabla}\right){\bf n}=0$ \cite{columns}.  Expanding this equation to quadratic order in $\delta{\bf n}\approx {\bf n} - \hat{\bf y}$ precisely yields Burgers equation in $v=-\hat{\bf x}\cdot\delta{\bf n}$.

\begin{figure}
\centering
\includegraphics[width=5cm]{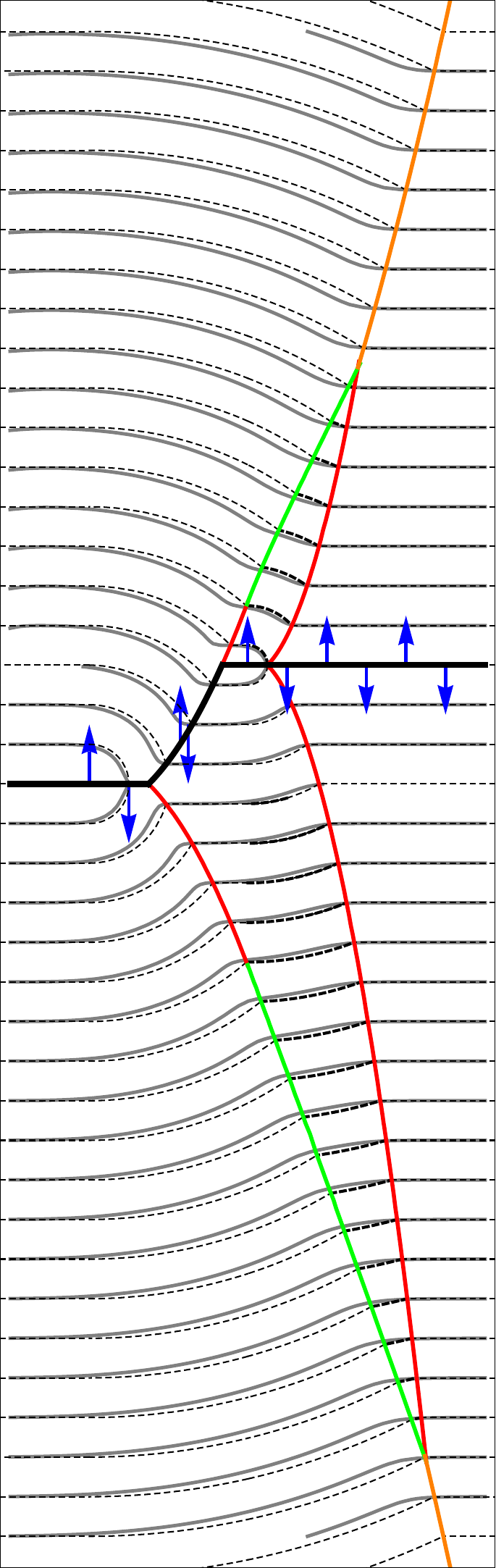}
\caption{(Color online) Two edge dislocations at arbitrary $(x_i,y_i)$. The dashed curves are the focal layers, and the solid, gray curves are BPS evolution with $\lambda=0.05$. Parabolic (red and orange) and hyperbolic (green) cusps of the focal construction are also shown. The thick solid lines indicates the division between upward and downward evolving BPS solutions. \label{2odisBPS}}
\end{figure}

When multiple defects lie along a line of constant $y$, the BPS method allows the superposition of defects via the Hopf-Cole transformation.  We can superpose in the focal construction too: multiple edge dislocations can be constructed by repeating the procedure described for Fig. \ref{1disfocal}. For example, in Fig. \ref{fig2G} we show the construction for a pair, both located at the same value of $y$. Note that there are now new features: in addition to parabolic focal curves, there are regions of the surface where cones intersect cones and, by definition, this happens along {\sl hyperbol\ae}. As we show in Fig. \ref{2odisBPS}, it is also possible to construct arbitrary focal textures in which the dislocations no longer lie at the same value of $y$: when lines meet circles they intersect on parabol\ae, when circles meet circles they intersect on a hyperbola.  

How does the BPS solution fare?  Again we begin with the deformations for large $\lambda/y$, where straightforward numerical analysis shows that hyperbol\ae\ are in the strain field 
\begin{eqnarray}\label{eq:2bps}
S &=&  1+\left(\text{e}^{b_1/4 \lambda}-1\right){\rm E}\left(\frac{x-x_1}{2\sqrt{\lambda y}}\right) \nonumber\\
&&\quad + \text{e}^{b_1/4 \lambda} \left(\text{e}^{b_2/4 \lambda} - 1\right){\rm E}\left(\frac{x-x_2}{2\sqrt{\lambda y}}\right) ,
\end{eqnarray}
corresponding to a pair of dislocations \cite{santangelo03}. Indeed, Fig. \ref{fig2G} shows remarkably good agreement between the focal construction and the BPS solution, including the details of the hyperbol\ae\ and the merging of the two focal curves. 

We also compute `exact' solutions for the level sets $\phi(x,y) = y - u(x,y)$, shown as dark solid and dashed lines in Fig. \ref{fig2G}, where the initial condition $u(x,0)$ is given by the phase field at $y=0$ in the focal construction. As expected from the asymptotic solution, the BPS evolution respects the parabolic cusps in the focal construction, deforming most to the left of the cusps but not on the right. 
This is to be expected; the deformation preferentially smooths out the higher curvature side and spreads the strain `inside' the parabolic region in agreement with the predictions of linear elasticity.
In BPS evolution, the quantities $S_\pm=\exp\{\pm u/(2\lambda)\}$ satisfy the extremal equations $ \partial_y S_\pm = \pm \lambda \partial_x^2 S_\pm$ \cite{santangelo03,newell}. The evolution has an inherent directionality: BPS evolution relaxes $S_+$ to flat layers above the dislocation and $S_-$ below the dislocation. Therefore, a dislocation at $y=0$ requires the BPS evolution to change directionality on either side of the line at $y=0$. Similarly, it is possible to find the textures generated by multiple dislocations, as long as they lie along the $y$-axis. 

When defects sit at different values of $y$, we have to be more careful when $\lambda>0$. It is instructive to consider the difficulty in detail. First, consider the focal construction shown in Fig. \ref{2odisBPS} (dashed lines). In the vicinity of each dislocation, we expect the solutions at finite $\lambda$ to be approximated by BPS evolution. Above and below both dislocations, there is no difficulty constructing a valid BPS evolution since the BPS evolution directions agree. The layers between the two dislocations, however, must evolve upward on the left and downward on the right. We can reconcile this discrepancy by noting that the parabolic cusp between dislocations in the focal texture also forms a natural division between upward and downward evolution. As shown in Fig. (\ref{2odisBPS}), we evolve upward on the left of the parabolic cusp using the displacement for the lower dislocation as the initial condition. On the right, we evolve downward using the phase field for the upper dislocation as an initial condition. 
The result of evolving upward and downward as indicated by the arrows in Fig. (\ref{2odisBPS}) is shown as solid layers. The layers arising from BPS evolution of opposite directionality meet naturally at the parabolic cusps {\sl without further} adjustment because the deformation field is strongly asymmetric, in this case confined to the left of the parabolic cusp. Were this not to occur, we could, of course, impose continuity of the layers at the cusp by setting the displacement of the upward evolution equal to that of the downward evolution. The success of the focal method hinges on the asymmetry of the distortion field for small $\lambda$.  Once we have constructed the shape of the layer on either side of the two dislocations, we may continue the evolution out to infinity. Again, the BPS evolution preserves the underlying structure of the cusps of the focal textures and the regions of maximum strain (and layer deviation) occur just to the left of the cusps.  As long as the defects are further apart than $\lambda$, this procedure should be reliable.

In summary, we have developed a focal construction for multiple (and arbitrary) configurations of dislocations in a smectic. This construction uncovers a deep relationship between the BPS evolution of single and multiple dislocations and the focal construction. Using the naturally occurring cusps in the focal construction, we are able to develop BPS solutions for dislocations with layers between them that account for the geometric nonlinearities in the elastic strain.

%%% acknowledgements %%%
It is a pleasure to acknowledge discussions with D. Beller, B.G. Chen, R. Kusner, E.A. Matsumoto, and R.A. Mosna. GPA and RDK were supported in part by NSF Grant DMR05-47230. CDS was supported in part by NSF Grant DMR08-46582.

\end{document}